\overfullrule=0pt
\input harvmac
\def\a{{\alpha}}
\def\l{{\lambda}}
\def\b{{\beta}}
\def\g{{\gamma}}
\def\d{{\delta}}
\def\e{{\epsilon}}

\def\O{{\Omega}}
\def\half{{1\over 2}}
\def\p{{\partial}}

\def\t{{\theta}}

\Title{\vbox{\hbox{IFT-P.005/2000 }}}
{\vbox{
\centerline{\bf Super-Poincar\'e Covariant Quantization of the Superstring}}}
\bigskip\centerline{Nathan Berkovits\foot{e-mail: nberkovi@ift.unesp.br}}
\bigskip
\centerline{\it Instituto de F\'\i sica Te\'orica, Universidade Estadual
Paulista}
\centerline{\it Rua Pamplona 145, 01405-900, S\~ao Paulo, SP, Brasil}

\vskip .3in
Using pure spinors, the
superstring is covariantly quantized. For the first time,
massless vertex operators are constructed and
scattering amplitudes are computed in a manifestly
ten-dimensional super-Poincar\'e covariant manner.
Quantizable non-linear sigma model actions are constructed for the
superstring in curved backgrounds, including the $AdS_5\times S^5$
background with Ramond-Ramond flux.

\Date {January 2000}

\newsec{Introduction}

There are many motivations for covariantly quantizing the superstring.
As in any theory, it is desirable to make all
physical symmetries manifest in order to reduce the amount of
calculations and simplify any cancellations coming from the symmetry.
Recently, an additional motivation has come from the desire to
construct a quantizable sigma model action for the superstring in 
curved backgrounds
with Ramond-Ramond flux.

Most attempts
to covariantly quantize the superstring have started from
the classical super-Poincar\'e invariant
version of the Green-Schwarz (GS) action \ref\GS{M.B. Green and J.H. Schwarz,
{\it Covariant Description of Superstrings}, Phys. Lett. B136 (1984) 367.}.
One quantization approach is based on gauge-fixing
the fermionic symmetries to get to
``semi-light-cone'' gauge where $(\gamma^+ \theta)_\alpha =0$ and
$\gamma^+ =\gamma^0 +\gamma^9$ \ref\carlip{S. Carlip,
{\it Heterotic String Path Integrals with the Green-Schwarz
Covariant Action}, Nucl. Phys. B284 (1987) 365
\semi R. Kallosh,
{\it Quantization of Green-Schwarz Superstring}, Phys. Lett. B195 (1987) 369.}.
In this gauge, the covariant Green-Schwarz action simplifies to
$S= \int d^2 z [ \p x^m 
\overline\p x_m + \p x^+(\theta \gamma^- \overline\p\theta)].$
However, even this simplified action cannot be easily quantized since
the propagator for $\theta$ involves $(\p x^+)^{-1}$ which is not
well-defined.\foot{On a genus $g$
worldsheet with $N$ punctures, $\p x^+$ vanishes at $2g+N-2$ points
on the worldsheet.
This fact is related to the need for interaction-point operators in
the light-cone GS superstring.}
For this reason, it has not
yet been possible to use this approach to construct
physical vertex operators or compute scattering
amplitudes, except in the $p^+\to 0$ limit that reproduces
the light-cone gauge computations \ref\john{G. Gilbert and
D. Johnston, {\it Equivalence of the Kallosh and Carlip Quantizations
of the Green-Schwarz Action for the Heterotic String}, Phys. Lett. B205
(1988) 273.}.
Another approach to quantizing the covariant
Green-Schwarz action is based on
replacing the fermionic second-class constraints with
an appropriate set of first-class constraints
\ref\csm
{W. Siegel, {\it Classical Superstring Mechanics}, Nucl. Phys. B263 (1986)
93.}, sometimes using 
SO(9,1)/SO(8) harmonic variables \ref\sok{E. Sokatchev, {\it
Harmonic Superparticle}, Class. Quant. Grav. 4 (1987) 237
\semi
E.R. Nissimov and S.J. Pacheva, {\it Manifestly Super-Poincar\'e
Covariant Quantization of the Green-Schwarz Superstring},
Phys. Lett. B202 (1988) 325\semi
R. Kallosh and M. Rakhmanov, {\it Covariant Quantization of the
Green-Schwarz Superstring}, Phys. Lett. B209 (1988) 233.} 
which covariantize the semi-light-cone gauge choice.
However, despite numerous attempts \ref\many{S.J. Gates Jr, M.T. Grisaru,
U. Lindstrom, M. Rocek, W. Siegel, P. van Nieuwenhuizen and
A.E. van de Ven, {\it Lorentz-Covariant Quantization of the Heterotic
Superstring}, Phys. Lett. B225 (1989) 44\semi
R.E. Kallosh, {\it Covariant Quantization of Type IIA,B
Green-Schwarz Superstring}, Phys. Lett. B225 (1989) 49\semi
M.B. Green and C.M. Hull, {\it Covariant Quantum Mechanics of the
Superstring}, Phys. Lett. B225 (1989) 57.}, noone was able to find an
appropriate set of first-class constraints which allows
the covariant computation of scattering amplitudes.

In the absence of Ramond states, it is possible to quantize the
superstring in a manifestly Lorentz-covariant manner using the standard 
Ramond-Neveu-Schwarz (RNS) formalism. However,
none of the spacetime supersymmetries
are manifest in the RNS formalism and, in order to explicitly 
construct the spin field for Ramond states,
manifest SO(9,1) Lorentz invariance must be broken (after Wick-rotation)
to a U(5) subgroup \ref\fms{D. Friedan, E. Martinec and S. Shenker,
{\it Conformal Invariance, Supersymmetry and String Theory},
Nucl. Phys. B271 (1986) 93.}. 
Recently, an alternative formalism for the superstring was
constructed which manifestly preserves this same U(5) subgroup
in addition to manifestly preserving six of the
sixteen spacetime supersymmetries \ref\ufive{N. Berkovits, {\it
Quantization of the Superstring with Manifest U(5) Super-Poincar\'e
Invariance}, Phys. Lett. B457 (1999) 94, hep-th/9902099.}. 
The worldsheet variables of this supersymmetric
U(5) formalism are
related to those of the RNS formalism by a field redefinition, allowing
one to prove that 
physical vertex operators and 
scattering amplitudes in the two formalisms
are equivalent. However, the lack of manifest Lorentz invariance
makes it difficult to use this formalism to describe the
superstring in curved (Wick-rotated)
backgrounds which do not preserve U(5) holonomy.

In this paper, a new formalism for the superstring will be presented
which can be quantized in a manifestly super-Poincar\'e covariant
manner. The worldsheet variables of this formalism will consist of the
usual ten-dimensional
superspace variables in addition to
a bosonic spacetime spinor $\lambda^\a$ satisfying the `pure'
spinor condition 
\eqn\pure{\l^\a \gamma^m_{\a\b} \l^\b=0}
for $m=0$ to 9. 
$\lambda^\a$ must be complex to satisfy \pure\ and,
after Wick-rotating SO(9,1) to SO(10), can be parameterized by
eleven complex variables. One of these eleven variables is an overall
scale factor, and the other ten parameterize the coset space
SO(10)/U(5). So this new formalism is probably
related to a covariantization of the U(5) formalism of \ufive. Although the
precise relation between the two formalisms is still unclear, it will
be argued in section 2 that pure spinor variables are necessary for
equating RNS vertex operators with the GS vertex operators proposed in \csm. 

In section 3, physical states will be defined as elements in the cohomology
of the BRST-like operator
\eqn\BRST{Q=\oint dz \l^\a d_\a}
where $d_\a$ is the generator of supersymmetric derivatives as 
defined in \csm.
Since $d_\a(y)d_\b(z)\to 2 (y-z)^{-1}\Pi_m(z)\g^m_{\a\b}$ 
where $\Pi_m$ is the supersymmetric
translation generator,
\pure\ implies that $Q^2=0$.
Note that the operator of \BRST\ was used in 
\ref\howe{P.S. Howe, {\it Pure Spinor Lines in Superspace and
Ten-Dimensional Supersymmetric Theories}, Phys. Lett. B258 (1991) 141,
Addendum-ibid.B259 (1991) 511\semi
P.S. Howe, {\it Pure Spinors, Function Superspaces and Supergravity
Theories in Ten Dimensions and Eleven Dimensions}, Phys. Lett. B273 (1991)
90.} by Howe to show that the constraints
of ten-dimensional
super-Yang-Mills and supergravity can be understood as integrability
conditions on pure spinor lines.

Using this definition of physical states, one can easily construct
the physical massless
vertex operators. For the open superstring, the massless vertex operator in
unintegrated form is $V = \lambda^\a A_\a(x,\t)$ and in integrated form is
\eqn\vertex{
V=\int dz (\Pi^m A_m + \p \t^\a A_\a + d_\a W^\a + N^{mn} F_{mn})}
where $A_M$ are the super-Yang-Mills prepotentials, $W^\a$ and $F_{mn}$
are the gauge-invariant superfields whose lowest components are the
gluino and the gluon field strength, and $N^{mn}$ is the
pure spinor contribution to the Lorentz generator. Except for the
$N^{mn}$ term, the vertex operator of \vertex\ is 
that proposed by Siegel in \csm. As will be shown in section 4,
these vertex operators can be used to compute scattering amplitudes
in a manifestly super-Poincar\'e covariant manner.

The physical vertex operators for the closed superstring can be 
obtained by taking the `left-right' product of two open superstring
vertex operators. In section 5, the integrated form of the closed
superstring massless vertex operator will be used to construct a quantizable
sigma model action
for the superstring in a curved superspace background.
As a special case, a quantizable sigma model action will be
constructed for the superstring in an $AdS_5\times S^5$ background with
Ramond-Ramond flux. This action differs from that of 
Metsaev and Tseytlin \ref\Metsaev{R. Metsaev and A. Tseytlin,
{\it Type IIB superstring action in $AdS_5 \times S^5$
background},
Nucl.Phys. {B533} (1998) 109, hep-th/9805028.}
in containing a kinetic term for the
fermions which allows quantization.

In section 6, further evidence will be given for equivalence with
the RNS formalism and some possible applications of the new formalism
will be discussed.

\newsec{Pure Spinors and Lorentz Currents}

In conformal gauge, the left-moving contribution to the 
covariant Green-Schwarz superstring action
can be written as
$$S=\int d^2 z (\half\p x^m \overline\p x_m + p_\a \overline\p \t^\a )$$
where $p_\a$ is related to $x^m$ and $\t^\a$ by the constraint
$d_\a=0$ with\csm
\eqn\dd{d_\a = p_\a +\g^m_{\a\b}\p x_m \t^\b +{1\over 2}\g^m_{\a\b}\g_{m\,\g\d}
\t^\b\t^\g\p\t^\d.}
Since $d_\a(y)d_\b(z)\to 2 (y-z)^{-1} \g^m_{\a\b} \Pi_m(z)$ where
$\Pi^m = \p x^m + \t^\a \g^m_{\a\b} \p\t^\b$,
$d_\a=0$ involves first and second-class constraints. 
The idea of \csm\ is to find an appropriate set of first-class constraints
constructed from $d_\a$
which can replace the second-class constraints. In such a framework,
$p_\a$ is treated as an independent field and physical vertex operators
are annihilated by the first-class constraints.
Although an appropriate set of first-class constraints were not found
in \csm, Siegel used supersymmetry arguments to conjecture 
that the massless open superstring vertex operator should have the form
\eqn\siegv{V=\int dz (\Pi^m A_m + \p \t^\a A_\a + d_\a W^\a)}
where $A_M$ are the super-Yang-Mills prepotentials and $W^\a$
is the super-Yang-Mills spinor field strength.

For a gluon, the vertex operator
of \siegv\ reduces to $V = \int dz (\p x^m {\cal A}_m(x) + \half
(p\g^{mn}\t) {\cal F}_{mn}(x))$
where ${\cal A}_m$ and ${\cal F}_{mn}$ are the ordinary $\t$-independent
gluon gauge field and field strength,
which closely resembles the gluon vertex operator in the RNS formalism
$V = \int dz (\p x^m {\cal A}_m + 
\psi^m\psi^n {\cal F}_{mn})$. However, there is
a crucial difference between the OPE's of the SO(9,1) Lorentz currents
$M^{mn}=
\half p\g^{mn}\t$  and 
$\widehat M^{mn}=
\psi^m\psi^n$
which will
force the introduction of pure spinors. Namely, the
OPE of 
$M^{kl}$ with $M^{mn}$ has a 
double pole proportional to ${{16}\over 4}(\eta^{kn}\eta^{lm} -
\eta^{km}\eta^{ln})$ where the factor of 16 comes from the spinor dimension.
However, the double pole in the OPE of $\widehat M^{kl}$ with 
$\widehat M^{mn}$ is proportional to $
(\eta^{kn}\eta^{lm} -
\eta^{km}\eta^{ln})$ without the factor of ${{16}\over 4}$.
So the vertex operator of \csm\ can only be equivalent at the
quantum level to the RNS vertex operator if one adds a new term to the 
Lorentz current $M^{mn} =
\half p\g^{mn}\t + N^{mn}$ where $N^{mn}$ satisfies the OPE
\foot{In four dimensions, $\half p\g^{mn}\t$
has a double pole proportional to 
${{4}\over 4}(\eta^{kn}\eta^{lm} -
\eta^{km}\eta^{ln})$, so there is no need to
add new Lorentz degrees of freedom when quantizing the four-dimensional
superstring \ref\four{N. Berkovits, 
{\it Covariant Quantization Of
The Green-Schwarz Superstring in a Calabi-Yau Background},
Nucl. Phys. B431 (1994) 258\semi
N. Berkovits, {\it A New Description Of The Superstring}, Proceedings to
VIII Jorge Swieca Summer School on Particles and Fields,
p. 490, World Scientific Publishing, 1996, hep-th/9604123.}.
In six dimensions, $\half  p_j\g^{mn}\t^j$ (where $j=1$ to 2 is an internal
SU(2) index)
has a double pole proportional to 
${{8}\over 4}(\eta^{kn}\eta^{lm} -
\eta^{km}\eta^{ln})$, so one needs to add degrees of freedom
whose Lorentz current has a double pole with itself
proportional to
$-(\eta^{kn}\eta^{lm} -
\eta^{km}\eta^{ln})$. These degrees of freedom are a bosonic spinor
$u^\a$ and its conjugate momentum $v_\a$ for $\a=1$ to 4.
They are the ghosts for the 
`harmonic' constraints $\widetilde d_\a= d_{\a 2}-e^{-\rho-i\sigma} d_{\a 1}$ 
of \ref\six{N. Berkovits,
{\it Quantization of the Type II Superstring in a Curved
Six-Dimensional Background}, to appear in Nucl. Phys. B, hep-th/9908041.}
whose contribution was incorrectly ignored in \six. The
correct massless six-dimensional open superstring vertex operator 
is 
$V=\int dz (\Pi^m A_m + \p \t^{\a j} A_{\a j} + d_{\a j} W^{\a j}
+ \half
(u\g^{mn} v) F_{mn})$ where $F_{mn}$ is a superfield whose lowest
component is the gluon field strength. Note that this vertex operator is
annihilated on-shell by the `harmonic' BRST-like operator $Q=\oint dz
u^\a \widetilde d_\a$
and the central charge contribution from $v_\a\p u^\a$
cancels the contribution from $p_{\a 2}\p\t^{\a 2}$ in the stress tensor
to give a vanishing conformal anomaly.}
\eqn\nope{N^{kl}(y) N^{mn}(z) \to
{{\eta^{m[l} N^{k]n}(z) - 
\eta^{n[l} N^{k]m}(z) }\over {y-z}} - 3
{{\eta^{kn} \eta^{lm} -
\eta^{km} \eta^{ln}}\over{(y-z)^2}}  .}

As will now be shown, such
a Lorentz current $N^{mn}$ can be explicitly constructed from 
a pure spinor $\l^\a$, i.e. a complex bosonic spinor
satisfying \pure. 
To parameterize the eleven
independent complex degrees of freedom
of $\l^\a$, it is convenient to Wick-rotate and
temporarily break SO(10) to SU(5)$\times$ U(1) as in \ufive. 
The sixteen complex
components of $\l^\a$ split into $(\lambda^+,\lambda_{ab}, \lambda^a)$
for $a,b=1$ to 5, 
which transform respectively
as $(1_{5\over 2},\overline{10}_{\half},5_{-{3\over 2}})$ 
representations of SU(5)$\times$ U(1) 
where the subscript denotes the U(1) charge.
In terms of the eleven independent
complex variables $(\g,u_{ab})$ transforming as
$(1_5, \overline{10}_{-2})$ representations, 
one can check that 
\eqn\lam{\lambda^+ = \g
, \quad \lambda_{ab} = \g u_{ab},\quad
\lambda^a =  -{1\over 8}\g\e^{abcde} u_{bc} u_{de}}
satisfies the pure spinor condition of \pure.
Note that $\g$-matrices in U(5) notation satisfy
$\l\g^a\l = \l^+\l^a +{1\over 8}\e^{abcde}\l_{bc}\l_{de}$ and
$\l\g_a\l = \l_{ab}\l^b$
where the SO(10) vector has been split into a $5_{1}$ and $\overline 5_{-1}$
representation.

In conformal gauge, the worldsheet action for the left-moving variables 
will be defined as 
\eqn\waction{S = \int d^2 z 
(\half\p x^m \overline\p x_m + p_\a \overline\p \t^\a  +
\half v^{ab}\overline\p u_{ab} + \beta\overline\p \gamma)}
with the left-moving stress tensor 
\eqn\stress{T = 
\half\p x^m \p x_m + p_\a \p \t^\a  +
\half v^{ab}\p u_{ab} + \beta\p \gamma}
where $(\beta,v^{ab})$ are the conjugate momenta for $(\gamma,u_{ab})$.
As desired, $T$ has no conformal anomaly since
the central contribution for the new degrees of freedom
is $22$, which cancels the central charge contribution from the $x^m$ and
$(\t^\a,p_\a)$ variables.

In U(5) notation, the SO(10) Lorentz currents $N^{mn}$ split into
$(N, N_a^b, N^{ab}, N_{ab})$ which transform respectively as
$(1_0, 24_0, 10_2, \overline{10}_{-2})$ representations.
After fermionizing $\gamma= \eta e^\phi$ and 
$\beta = \p\xi e^{-\phi}$ as in \fms, $N^{mn}$ will be defined as  
\eqn\lorentz{N = {1\over{\sqrt{5}}}( u_{ab} v^{ab} + {{25}\over 4}\eta\xi
+{{15}\over 4} \p\phi), \quad 
N_a^b =  u_{ac} v^{bc} -{1\over 5}\d_a^b u_{cd}v^{cd}, }
$$N^{ab} = v^{ab}, \quad 
N_{ab} = 3\p u_{ab} + u_{ac} u_{bd} v^{cd} + 
u_{ab} ({5\over 2}\eta\xi +{3\over 2}\p\phi).$$ 
Using the free-field OPE's, 
\eqn\fope{\eta(y)\xi(z) \to (y-z)^{-1},\quad
\phi(y)\phi(z) \to -\log(y-z),\quad v^{ab}(y) u_{cd}(z) \to
\d^{[a}_c \d^{b]}_d (y-z)^{-1},}
one can check that 
\eqn\check{N_a^b(y) N_c^d(z) \to {{\d^b_c N^d_a (z) -\d^d_a N^b_c}\over
{y-z}} -3 {{\d_a^d \d_c^b -{1\over 5}\d_a^b\d_c^d}\over{(y-z)^2}},\quad
N(y)N(z)\to -{3\over{(y-z)^2}},}
$$ N_{ab}(y) N^{cd}(z)
\to {{-\d_{[a}^{[c} N_{b]}^{d]}(z) -{2\over{\sqrt{5}}}
\d_{a}^{[c}\d_{b}^{d]}N(z)}
\over{y-z}} +3
{{\d_{a}^{[c}\d_{b}^{d]}} \over{(y-z)^2}},$$
which correctly reproduces the OPE of \nope.

Furthermore, $[\oint dz N^{mn},\l^a] = \half(\g^{mn}\l)^\a$ as can
be easily shown
by noting that
$N^{mn} = \half\omega \gamma^{mn}\l$
where $\omega_\a$ is a spinor of the opposite chirality to $\l^\a$ with
components\foot
{Terms coming from normal-ordering ambiguities in $\omega \g^{mn}\l$
can be ignored since they only involve 
$\p\phi+\eta\xi$ and $\p u_{ab}$, which have no singularities with $\l^\a$. }
\eqn\wcomp{\omega_+ = \xi e^{-\phi} (\eta\xi -\half u_{ab}v^{ab}),\quad
\omega^{ab} = \xi e^{-\phi} v^{ab},\quad \omega_a =0.}
Note that 
\eqn\wope{\omega_\a(y)\l^\b(z) \to (y-z)^{-1}
\d_\a^\b -\half (y-z)^{-1} \g_m^{\b +} \xi e^{-\phi} (\g^m\l)_\a}
where the $+$ in $\g_m^{\b +}$ signifies the $1_{{5\over 2}}$
spinor component in the SU(5) notation of \lam. The second term
in the OPE of \wope\ is necessary for $\omega$ to have no singularity
with $\l\g^m\l$, however, it does not contribute to the commutator
$[\oint dz\omega \gamma^{mn}\l, \l]$
since $\l \g^m\g^{np}\l=0.$ 

So after introducing pure spinors, it is possible to obtain vanishing
conformal anomaly and to relate the RNS gluon vertex operator with the
proposal of Siegel in \csm. It will now be shown how these pure spinors
can be used to define physical vertex operators and compute
scattering amplitudes in a super-Poincar\'e covariant manner.

\newsec{Physical Vertex Operators}

Since the stress-tensor of \stress\ has vanishing central charge, 
one can require that physical vertex operators in unintegrated form
are primary fields of dimension zero. However, this requirement is
clearly insufficient since, for a massless vertex operator depending
only on the zero modes of the worldsheet fields, it implies
$\p_m\p^m \Phi(x,\t,\l)=0$ which 
has far more propagating fields than super-Yang-Mills. One
therefore needs a further constraint on physical vertex operators,
and using the intuition of \csm, this constraint should be
constructed from $d_\a$ of \dd. 

Using the pure spinor $\l^\a$ defined in terms of $\g$ and $u_{ab}$ as
in \lam, one can define a nilpotent BRST-like operator
\eqn\brstlike{Q = \oint dz\l^\a(z) d_\a(z).}
Defining ghost charge to be $q_{ghost}=\oint dz \gamma\b$, $Q$ carries
ghost-number one. So it is natural to define physical vertex operators
as states of ghost-number 1 in the cohomology of $Q$. Note that
after Wick rotation, $\t^\a$ and $\l^\a$ are complex spinors, so
the Hilbert space of states should be restricted to analytic functions
of these variables.\foot{Although it is difficult to impose reality
conditions on the states in Euclidean space, this is not a problem
for computing scattering amplitudes since it will be trivial to Wick-rotate
the final result back to Minkowski space where the reality conditions
are easily defined.}

It will now be shown for the massless sector of the open superstring that
this definition of physical states reproduces the desired super-Yang-Mills
spectrum. Massless vertex operators of dimension zero
can only depend on the worldsheet
zero modes, so the most general such vertex operator of ghost number 1 is
$U= \g Y(x,\t,u_{ab})$ where $Y$ is an analytic function of $\t^\a$ and
$u_{ab}$. Since $Q$ is Lorentz invariant (after including the contribution
of $N^{mn}$ \lorentz\ in the Lorentz generators),
elements in its cohomology must transform Lorentz covariantly. But
because of the non-linear nature of the Lorentz transformations generated
by $N^{mn}$, the only finite-dimensional covariantly transforming object
which is linear in $\g$ is $\l^\a$. So if the cohomology is restricted
to finite-dimensional elements, the most general massless vertex operator
of ghost number 1 is
\eqn\unint{U = \l^\a A_\a(x,\t)}
where $A_\a(x,\t)$ is a generic spinor function of $x^m$ and $\t^\a$.

The constraint $QU=0$ implies that $\l^\a \l^\b D_\b A_\a =0$
where $D_\a = {\p\over{\p\t^\a}} +\g^m_{\a\b}\t^\b {\p\over{\p x^m}}$.
Since $\l\g^m\l=0$, this implies that $D_\a (\g^{mnpqr})^{\a\b} A_\b =0$,
which is the on-shell constraint for the spinor prepotential of
super-Yang-Mills. Furthermore, the gauge transformation 
\eqn\gauge{\d U = Q \Omega(x,\t) = \l^\a D_\a \Omega(x,\t)}
reproduces the usual super-Yang-Mills gauge transformation
$\d A_\a = D_\a \Omega$ where $\Omega(x,\t)$ is a generic scalar
superfield. So the ghost number 1 cohomology of $Q$
for the massless sector reproduces the desired super-Yang-Mills
spectrum. 

To compute scattering amplitudes, one also needs vertex operators
in integrated form, i.e. integrals of dimension 1 primary fields.
Normally, these are obtained from the unintegrated vertex operator
by anti-commuting with $b(z)$. But in this formalism, there
are no $(b,c)$ ghosts, so it is presently unclear how to
relate the two types of vertex operators. Nevertheless, one can
define physical integrated vertex operators as elements
in the BRST cohomology of ghost-number zero.

In the massless sector, there is an obvious candidate which
is the dimension 1 vertex operator of \siegv\ suitably modified to include
the pure spinor contribution to the Lorentz current, i.e. 
\eqn\hatv{V = 
\Pi^m A_m + \p \t^\a A_\a + d_\a W^\a +  N^{mn}F_{mn}}
where $F_{mn}$ is the superfield whose lowest component is
the gluon field strength. To show that $[Q ,\int dz V]=0$,
first note that\csm
\eqn\openote{d_\a(y) \int dz v(z) \to \half \int dz 
(y-z)^{-1} F_{mn}(z) (d(z)\g^{mn})_\a}
where
$v(z)=
\Pi^m A_m + \p \t^\a A_\a + d_\a W^\a.$
Since
$ \l^\a(y) N^{mn}(z) \to \half (y-z)^{-1} (\l(z)\g^{mn})^\a$,
$$[Q , \int dz V]= \int dz  N^{mn} \l^\a D_\a F_{mn} = 
\int dz N^{mn} (\l \g_m \p_n W)$$
where $W^\a$ is the spinor field strength.
But using $N^{mn} = \half\omega\g^{mn}\l$ from \wcomp, 
$$N^{mn} (\l\g_m)_\a = \half(w\g^{mn}\l)(\l\g_m)_\a = \half
(w_\b\l^\b) (\l\g^n)_\a$$
since $(\g^m\l)_\a (\g_m\l)_\b =
-\half\g^m_{\a\b} (\l\g_m\l) =
0 $ from ten-dimensional
$\g$-matrix identities.
Finally, using the gluino equation of motion, 
$$[Q ,\int dz V]= \half\int dz  
(w_\b\l^\b) (\l\g^n \p_n W)=0$$
so $\int dz V$ describes a physical integrated vertex operator.
Note that the super-Yang-Mills gauge transformation $\d A_M = \p_M \Omega$
transforms $V$ by the total derivative $\p_z \Omega$, so
$\int dz V$ is manifestly gauge-invariant.

\newsec {Computation of Scattering Amplitudes}

In this section, it will be shown how to compute tree-level
open superstring scattering amplitudes 
in a manifestly super-Poincar\'e covariant manner.
To compute $N$-point tree-level scattering amplitudes, one needs three
vertex operators in unintegrated form 
and $N-3$ vertex operators in integrated form. Since only
the massless vertex operators are known explicitly, only scattering
of massless states will be considered here.

The two-dimensional
correlation function which needs to be evaluated for computing
tree-level scattering of $N$ super-Yang-Mills multiplets is
\eqn\pres{{\cal A} = \langle U_1(z_1) U_2(z_2) U_3(z_3)
\int dz_4 V_4(z_4) ...
\int dz_N V_N(z_N)\rangle}
where $U_r$ is the dimension 0 vertex operator of \unint,
$V_r$ is the dimension 1 vertex operator of \hatv, and the locations
of $(z_1,z_2,z_3)$ can be chosen arbitrarily because of SL(2,R) invariance. 

The functional integral over the non-zero modes of the various worldsheet
fields is completely straightforward using the free-field OPE's.
For example, the dimension 1 worldsheet fields $(\p x^m, \p\t^\a,
d_\a, N^{mn})$ can be integrated out by contracting with 
other dimension 1 fields or with $(x^m,\t^\a,\l^\a)$. Note
that manifest Lorentz invariance is preserved by the contractions
of $N^{mn}$ because its only singular OPE's are
$N^{mn}(y)\l^\a(z)\to \half(y-z)^{-1}(\g^{mn}\l)^\a$
and \nope.

However, the functional integral prescription for the zero modes
of the worldsheet fields needs to be explained. Besides the zero
modes of $x^m$ (which are treated in the usual manner using
conservation of momentum), there are the eleven bosonic zero modes
of $(\g,u_{ab})$ and the sixteen fermionic zero modes of $\t^\a$.
After integrating out the non-zero modes, one gets an expression
\eqn\zerop{{\cal A} = \int dz_4 ...\int dz_N
\langle \l^\a \l^\b \l^\g
f_{\a\b\g}(z_r, k_r, \t)\rangle}
where only the zero modes of $\l^\a$ contribute
and $f_{\a\b\g}$ is a function which depends on $z_4 ... z_N$,
on the momenta $k_r^m$ for $r=1$ to $N$, and on the zero modes of $\t^\a$.

The prescription for integration over the remaining worldsheet
zero modes will be 
\eqn\zeroq{\langle \l^\a \l^\b \l^\g
f_{\a\b\g}(z_r, k_r, \t)\rangle
\equiv  \int d\Omega (\overline\l_\d \l^\d)^{-3}
({\p\over{\p\t}}\g^{lmn}{\p\over{\p\t}}) (\overline\l\g_l{\p\over{\p\t}})
(\overline\l\g_m{\p\over{\p\t}}) 
(\overline\l\g_n{\p\over{\p\t}}) }
$$ \l^\a \l^\b\l^\g 
f_{\a\b\g}(z_r, k_r, \t)$$
where $\overline\l_\a$ is the complex conjugate of $\l^\a$
in Euclidean space
and $d\Omega$ is an integration over the different
possible orientions of $\l^\a$.
Although this prescription is defined in Euclidean space, it is
trivial to Wick-rotate the result back to Minkowski space using
the fact that 
\eqn\rota{\int d\Omega (\overline \l_\d \l^\d)^{-3} \l^\a \l^\b \l^\g
\overline\l_\rho \overline\l_\sigma 
\overline\l_\tau = T^{\a\b\g}_{\rho\sigma\tau}
\equiv {1\over{4032}}
[\d_\rho^{(\a} \d_\sigma^\b \d_\tau^{\g)} - {1\over{40}}
\g_m^{(\a\b} \d_{(\rho}^{\g)}\g^m_{\sigma\tau)}].}
Equation \rota\ can be derived using the fact that there is
a unique covariantly transforming tensor $T^{\a\b\g}_{\rho\sigma\tau}$ 
which is symmetrized with respect to its upper and lower indices and
which satisfies $T^{\a\b\g}_{\a\b\g}=1$ and 
$T^{\a\b\g}_{\rho\sigma\tau} \g^m_{\a\b} = 
T^{\a\b\g}_{\rho\sigma\tau} \g_m^{\rho\sigma} = 0.$

So the amplitude
of \zerop\ can be written in SO(9,1) Lorentz-covariant notation as
\eqn\mink{{\cal A} =  T^{\a\b\g}_{\rho\sigma\tau}
({\p\over{\p\t}}\g^{lmn}{\p\over{\p\t}}) (\g_l{\p\over{\p\t}})^\rho
(\g_m{\p\over{\p\t}})^\sigma 
(\g_n{\p\over{\p\t}})^\tau  \int dz_4 ... \int dz_N
f_{\a\b\g}(z_r, k_r, \t)}
where 
$T^{\a\b\g}_{\rho\sigma\tau}$ is defined in \rota.
By expanding $f_{\a\b\g}(z_r,k_r,\t)$ as a power series in $\t^\a$,
one can check that the prescription of \zeroq\ selects out the term
\eqn\selects{
f_{\a\b\g}(z_r,k_r,\t) = ... + a(z_r,k_r) 
(\t\g^{lmn}\t) (\g_l\t)_\a
(\g_m\t)_\b
(\g_n\t)_\g + ...}
in the power series, i.e. 
$\langle \l^\a \l^\b \l^\g
f_{\a\b\g}(z_r, k_r, \t)\rangle = a(z_r,k_r)$.

This prescription for integrating out the zero modes is reasonable
since it is Lorentz invariant and since the eleven bosonic zero mode
integrations
are expected to cancel eleven of the sixteen fermionic zero mode integrations,
leaving five zero modes of $\t$ which are removed with five 
${\p\over{\p\t}}$'s.
Further evidence for this zero-mode prescription comes from the fact
that it is gauge invariant and spacetime supersymmetric, as will now be shown.

To show that ${\cal A}$ is invariant under a gauge transformation
$\d U_1(z_1) 
= [\oint dz \l^\a d_\a, \Omega(z_1)]$, note that 
$\oint dz \l^\a d_\a$ commutes with $U_r$ and $\int dz_r V_r$, so
\eqn\transfg{\d{\cal A} = \langle [\oint dz\l^\a d_\a ,   \Omega(z_1)
 U_2(z_2) U_3(z_3)
\int dz_4 V_4(z_4) ...
\int dz_N V_N(z_N) ]\rangle}
$$ =\int dz_4 ... \int dz_N
 \langle \l^\a  \l^\b \l^\g D_\a h_{\b\g}(z_r,k_r,\t)\rangle$$
for some $h_{\b\g}$
after integrating out the non-zero modes.
Using the zero-mode prescription of \mink,
\eqn\minkdelta{\delta{\cal A} =  T^{\a\b\g}_{\rho\sigma\tau} 
({\p\over{\p\t}}\g^{lmn}{\p\over{\p\t}}) (\g_l{\p\over{\p\t}})^\rho
(\g_m{\p\over{\p\t}})^\sigma 
(\g_n{\p\over{\p\t}})^\tau {\p\over{\p\t^\a}}  \int dz_4 ... \int dz_N
h_{\b\g}(z_r, k_r, \t),}
where conservation of momentum has been used to replace $D_\a$ with
${\p\over{\p\t^\a}}$ in \minkdelta.
But using anti-symmetry properties of ${\p\over{\p\t}}$, one can show that
$$T^{\a\b\g}_{\rho\sigma\tau} 
({\p\over{\p\t}}\g^{lmn}{\p\over{\p\t}}) (\g_l{\p\over{\p\t}})^\rho
(\g_m{\p\over{\p\t}})^\sigma 
(\g_n{\p\over{\p\t}})^\tau {\p\over{\p\t^\a}} =0,$$
so $\d{\cal A}=0$.
 
It will now be shown that the prescription of \mink\ is invariant
under spacetime supersymmetry transformations, implying that the amplitudes
are SO(9,1) super-Poincar\'e invariant. 
Under a spacetime supersymmetry transformation with global parameter 
$\e^\kappa$, the term $a(z_r,k_r)$ of \selects\ transforms as
$\d a(z_r,k_r) = \e^\kappa \xi_{\kappa}(z_r,k_r)$ where 
$\xi_{\kappa}$ appears in the power series for $f_{\a\b\g}$ as
\eqn\xisel{
f_{\a\b\g}(z_r,k_r,\t) = ... + 
(\t\g^{lmn}\t) (\g_l\t)_\a
(\g_m\t)_\b
(\g_n\t)_\g  [ a(z_r,k_r) + \t^\kappa \xi_\kappa(z_r,k_r) ] + ... .}
But $\int dz_4 ... \int dz_N \l^\a\l^\b\l^\g f_{\a\b\g}$ comes
from vertex operators which commute with $\oint dz \l^\d d_\d$,
so it 
must satisfy the constraint 
\eqn\musts{\int dz_4 ... \int dz_N \l^\a\l^\b\l^\g \l^\d D_\d f_{\a\b\g}=0}
for any pure spinor $\l^\a$.
Plugging \xisel\ into \musts\ and using 
$(\l\g^{lmn}\t) (\l\g_l\t) (\l\g_{m}\t) (\l\g_n\t) =0$, one
finds that 
\eqn\mtwo{\int dz_4 ... \int dz_N 
(\t\g^{lmn}\t) (\l\g_l\t) (\l\g_{m}\t) (\l\g_n\t)\l^\kappa \xi_\kappa(z_r,k_r)}
must vanish for any pure spinor $\l^\a$. But this is only possible if
$\int  dz_4 ... \int dz_N  \xi_\kappa(z_r,k_r) =0$, 
implying that 
\eqn\implya{\d{\cal A} = \int dz_4 ... \int dz_N \d a(z_r,k_r) =0}
so the amplitude prescription of \mink\ is spacetime supersymmetric.

\newsec {Superstring Action in a Curved Background}

In this section, the massless integrated vertex operator
for the closed superstring
will be used to construct a quantizable action for the superstring
in a curved background. As a special case, a quantizable action will
be constructed for
the Type IIB
superstring in an $AdS_5\times S^5$ background with Ramond-Ramond flux.

In bosonic string theory and in the Neveu-Schwarz sector of superstring
theory, the action in a curved background (ignoring the
Fradkin-Tseytlin term for dilaton coupling) can be constructed by 
`covariantizing' the massless closed string vertex operator with
respect to target-space reparameterization invariance. As in 
\ref\eff{N. Berkovits and W. Siegel, {\it
Superspace Effective Actions for 4D Compactifications
of Heterotic and Type II Superstrings},
Nucl. Phys. B462 (1996) 213, hep-th 9510106.} and
\six, this procedure can also be used here 
after constructing the
massless closed string
vertex operator from the `left-right'
product of two massless open string vertex operators of \hatv. 

To do this,
one first needs to introduce right-moving analogs of the worldsheet fields
described in \waction. The complete worldsheet action for the
Type II superstring in a flat background
in conformal gauge is
\eqn\caction{
S = \int d^2 z 
(\half\p x^m 
\overline\p x_m + p_\a \overline\p \t^\a  + \widehat p_{\widehat\a}
\p \widehat\t^{\widehat\a} +
\half (v^{ab}\overline\p u_{ab} 
+ \widehat v^{ab}\p \widehat u_{ab} )
+ \beta\overline\p \gamma
+ \widehat\beta\p \widehat\gamma
)}
where $\widehat\l^{\widehat\a}$ 
is constructed from $\widehat\g$ and $\widehat u_{ab}$ in
a manner similar to \lam. Note that $\widehat \t^{\widehat\a}$ and 
$\widehat\l^{\widehat\a}$ are independent of $\t^\a$ and $\l^\a$,
and are not related by complex conjugation.
For the Type IIA superstring, the hatted spinor index has the opposite
chirality to the unhatted spinor index while, for the Type IIB superstring, the
hatted spinor index has the same chirality as the unhatted spinor index.

The action for the Type II superstring in a curved background obtained
by
`covariantizing' the massless closed superstring vertex operator 
with respect to target-space super-reparameterization invariance is\foot
{In the action for the
superstring in a curved six-dimensional background, there are terms
coming from the bosonic ghosts $(u^\a,v_\a)$ described in footnote 3 which were
incorrectly omitted from the action of \six.
The correct action should have terms of the type 
$\int d^2 z (d_{\a j} +  u_\a v^\b D_{\b j}) E^{\a j}_M\overline\p Y^M + ...)$,
as well as a kinetic action for the 
$(u^\a,v_\a)$ and $(\widehat u^{\widehat\a},
\widehat v_{\widehat\a})$ ghosts.}
\eqn\cvertex{S=\int d^2 z [\half
(G_{MN}+ B_{MN}) \p y^M \overline\p y^N }
$$+
 (d_\a +  N_\a^\b D_\b) E^\a_M\overline\p y^M
+ (\widehat d_{\widehat\a} + 
\widehat N_{\widehat\a}^{\widehat\b} \widehat D_{\widehat\b}) 
E^{\widehat\a}_M \p y^M
+ (d_{\a} +N_\a^\b D_\b) 
(\widehat d_{\widehat\g}+ \widehat N_{\widehat\g}^{\widehat\d}
 \widehat D_{\widehat\d})
P^{\a \widehat\g } $$
$$
+ 
\half (v^{ab}\overline\p u_{ab} 
+ \widehat v^{ab}\p \widehat u_{ab}) 
+ \beta\overline\p \gamma
+ \widehat\beta\p \widehat\gamma ]$$
where $y^M$ parameterizes the curved superspace background,
$E_M^\a$ and $E_M^{\widehat\a}$ 
are the spinor parts of the super-vierbein $E^A_M$,
$N_\a^\b \equiv N^{mn} (\g_{mn})_\a^\b$ 
and $\widehat N_{\widehat\a}^{\widehat\b} \equiv\widehat N^{mn} 
(\g_{mn})_{\widehat\a}^{\widehat\b}$ 
with 
$N_{mn}$ being defined by \lorentz\ and $\widehat N_{mn}$ being defined
similarly in terms of the hatted variables, 
and
$P^{\a \widehat\g}$ is the superfield whose
lowest components are the bispinor Ramond-Ramond field strengths.
The operators $D_\a$ and $\widehat D_{\widehat\a}$ in 
\cvertex\ are understood to act on the superfield to their left, e.g.
on $E^\a_M$, $E^{\widehat\a}_M$ or $P^{\a\widehat\g}$.

Note that the first line
of \cvertex\ is identical to the Green-Schwarz action in a curved 
background, however, the second and third lines are crucial
for quantization since they provide an invertible propagator for
$\t^\a$ and $\widehat\t^{\widehat\a}$.
Furthermore, since there is no fermionic $\kappa$-symmetry which
needs to be preserved, 
there is no problem with adding a Fradkin-Tseytlin
term to \cvertex\ of the type $\a'
\int d^2 z \Phi(x,\t,\widehat\t) R$ 
where $R$ is the worldsheet curvature and $\Phi$ is a scalar superfield
whose lowest component is the dilaton.

When the background superfields satisfy their effective low-energy
equations of motion, the action of 
\cvertex\ together with the Fradkin-Tseytlin term
is expected to be conformally invariant where the left-moving
stress tensor is
\eqn\stresscurv{
T=\half
G_{MN}\p y^M \p y^N 
+
 (d_\a +  N_\a^\b D_\b) E^\a_M\p y^M
+ 
\half v^{ab}\p u_{ab} 
+ \beta\p \gamma + \a' \p^2 \Phi .}
Furthermore, the current $\l^\a d_\a$ is expected to be
holomorphic and nilpotent when the background superfields are on-shell.

\subsec{Superstring Action in $AdS_5\times S^5$ background}

In this subsection, the action of \cvertex\ will be
explicitly constructed for the special case of the Type IIB superstring in
an $AdS_5\times S^5$ background with Ramond-Ramond flux.
As discussed in \Metsaev, this background
can be conveniently described by a coset supergroup $g$ taking values
in $PSU(2,2|4)/SO(4,1)\times SO(5)$ where the super-vierbein $E^A_M$ satisfies
\eqn\superv{ E^A_M dy^M =(g^{-1} dg )^A}
and $A=(c,\a,\widehat\a)$ 
ranges over the 10 bosonic and 32 fermionic entries in the
Lie-algebra valued matrix $g^{-1}dg$.
Furthermore, as discussed in \ref\bersh{N. Berkovits, M. Bershadsky, 
T. Hauer, S. Zhukov and B. Zwiebach, {\it Superstring Theory on
$AdS_2\times S^2$ as a Coset Supermanifold}, to appear in Nucl. Phys. B,
hep-th/9907200.}, the only non-zero components of $B_{AB} = E_A^M E_B^N
B_{MN}$ and $P^{\a\widehat\b}$ are 
\eqn\further{ B_{\a\widehat\b} =  B_{\widehat\b \a} = 
-{1\over {2ng_s}} \d_{\a\widehat\b} ,\quad P^{\a\widehat\b}
 =ng_s \d^{\a\widehat\b},}
where $n$ is the value of the Ramond-Ramond flux, $g_s$ is the string
coupling constant, and $\d_{\a\widehat\b} = (\g^{01234})_{\a\widehat\b}$
with $01234$ being the directions of $AdS_5$.

Plugging these background superfields into the action of \cvertex, one
finds
\eqn\ads{S_{AdS}=\int d^2 z 
[\half \eta_{cd} (g^{-1} \p g)^c (g^{-1} \overline\p g)^d
-{1\over{4ng_s}}\d_{\a\widehat\b}
[ (g^{-1} \p g)^\a (g^{-1} \overline\p g)^{\widehat \b} -
(g^{-1} \overline\p g)^{\a} (g^{-1} \p g)^{\widehat\b} ]}
$$+
 d_\a  (g^{-1} \overline\p g)^\a
+ \widehat d_{\widehat\a}  (g^{-1}\p g)^{\widehat\a}
+ n g_s \d^{\a\widehat\b} d_{\a}\widehat d_{\widehat\b}
$$
$$
+ N_{cd} (g^{-1} \overline\p g)^{cd}
+ \widehat N_{cd} (g^{-1} \p g)^{cd} +
\half (v^{ab}\overline\p u_{ab} 
+ \widehat v^{ab}\p \widehat u_{ab}) 
+ \beta\overline\p \gamma
+ \widehat\beta\p \widehat\gamma ]$$
where $(g^{-1} dg)^{cd}$ are the $SO(4,1)\times SO(5)$
coset elements of $g^{-1} dg$. This action is $PSU(2,2|4)$-invariant
since under $g\to M g \Omega$ with $M$ a global $SU(2,2|4)$ matrix
and $\Omega$ a local $SO(4,1)\times SO(5)$ matrix assumed to be close
to the identity,
\eqn\trns{\d (g^{-1} dg)^A = ([ g^{-1} dg, \O])^A, \quad
\d (g^{-1} dg)^{cd} = ([g^{-1} dg, \O])^{cd} + (d\O)^{cd}, }
$$\d
(\half (v^{ab}\overline\p u_{ab} 
+ \widehat v^{ab}\p \widehat u_{ab}) 
+ \beta\overline\p \gamma
+ \widehat\beta\p \widehat\gamma ) = -N_{cd} (\overline\p \O)^{cd}
- \widehat N_{cd} (\p \O)^{cd}, $$
where $(d_\a,\widehat d_{\widehat\a})$ and
$(\l^\a,\widehat \l^{\widehat\a})$ 
are defined to transform as Lorentz-covariant spinors under the local
$SO(4,1)\times SO(5)$ transformation.

Because of the  $ n g_s \d^{\a\widehat\b} d_\a \widehat d_{\widehat \b}$
term, $d_\a$ and $\widehat d_{\widehat\a}$ are auxiliary fields which can
be integrated out as was done in \ref\wit{N. Berkovits, C. Vafa
and E. Witten, {\it Conformal Field Theory of AdS Background
with Ramond-Ramond Flux}, JHEP 9903 (1999) 018, hep-th/9902098.}.
Their auxiliary equations of motion are 
\eqn\aux{d_\a =  {1\over{ng_s}} \d_{\a\widehat\b} 
(g^{-1} \p g)^{\widehat\b},\quad
\widehat d_{\widehat\b} = -{1\over{ng_s}} 
\d_{\a\widehat\b} (g^{-1} \p g)^{\a},}
which can be substituted into \ads\ to give
\eqn\adt{S_{AdS}=\int d^2 z 
[\half \eta_{cd} (g^{-1} \p g)^c (g^{-1} \overline\p g)^d }
$$
-{3\over{4ng_s}}\d_{\a\widehat\b}
(g^{-1} \overline\p g)^{\a} (g^{-1} \p g)^{\widehat\b} 
-{1\over{4ng_s}}\d_{\a\widehat\b}
(g^{-1} \p g)^{\a} (g^{-1} \overline\p g)^{\widehat\b} $$
$$
+ N_{cd} (g^{-1} \overline\p g)^{cd}
+ \widehat N_{cd} (g^{-1} \p g)^{cd} +
\half (v^{ab}\overline\p u_{ab} 
+ \widehat v^{ab}\p \widehat u_{ab}) 
+ \beta\overline\p \gamma
+ \widehat\beta\p \widehat\gamma ].$$
Finally, one can perform the rescaling
\eqn\rescaling{E^c_M \to (ng_s)^{-1},\quad E^\a_M \to (ng_s)^{-\half},\quad
E^{\widehat\a}_M \to (ng_s)^{-\half},}
to obtain the action\foot{ Two quantizable actions have been proposed
for the superstring in an 
$AdS_3\times S^3$ background with Ramond-Ramond flux \wit\six. 
One of the actions
includes eight left and right-moving $\t$ coordinates and is based on
the coset supergroup $PSU(1,1|2)\times PSU(2|2)/SU(2)\times SU(2)$ \six.
For the reasons discussed in footnotes 3 and 6, this action requires
the addition of terms involving the bosonic ghosts $(u^\a,v_\a)$ and
$(\widehat u^{\widehat\a},
\widehat v_{\widehat\a})$.
The other action includes four left and right-moving $\t$ coordinates
and is based on the supergroup $PSU(2|2)$ \wit. This action
does not involve the `harmonic' constraints discussed in \six\ and
therefore does not require the addition of any terms involving 
bosonic ghosts.}
\eqn\adu{S_{AdS}={1\over{ n^2 g_s^2}}\int d^2 z 
[\half \eta_{cd} (g^{-1} \p g)^c (g^{-1} \overline\p g)^d }
$$-{3\over{4}}\d_{\a\widehat\b}
(g^{-1} \overline\p g)^{\a} (g^{-1} \p g)^{\widehat\b} 
-{1\over{4}}\d_{\a\widehat\b}
(g^{-1}\p g)^{\a} (g^{-1} \overline\p g)^{\widehat\b}] $$
$$
+ \int d^2 z [N_{cd} (g^{-1} \overline\p g)^{cd}
+ \widehat N_{cd} (g^{-1} \p g)^{cd} + 
\half (v^{ab}\overline\p u_{ab} 
+ \widehat v^{ab}\p \widehat u_{ab}) 
+ \beta\overline\p \gamma
+ \widehat\beta\p \widehat\gamma ].$$
Except for the third line of \adu\
involving the pure spinor fields, this is precisely
the action which was proposed in \bersh. Note that unlike the 
action proposed 
in \Metsaev, the action of \adu\ is straightforward to quantize
and sigma model loop computations can be performed in the manner of \bersh.

\newsec{Concluding Remarks}

In this paper, a new formalism has been presented for covariantly quantizing
the superstring. For the first time, vertex operators have been constructed
and scattering amplitudes have been computed in a manifestly
ten-dimensional super-Poincar\'e
invariant manner. A quantizable action has been proposed for the
superstring in any curved background, including the $AdS_5\times S^5$
background with Ramond-Ramond flux.

There are various possible generalizations of the new formalism which
should be possible. For example, one should be able to generalize the massless
vertex operators to massive vertex operators and generalize
the tree-level amplitude prescription
to a multiloop amplitude prescription. One should
also be able to construct physical vertex operators in the $AdS_5\times S^5$
background in a manner similar to the construction of
$AdS_3\times S^3$ vertex operators in
\ref\dolan{L. Dolan and E. Witten, {\it Vertex Operators
for $AdS_3$ Background with Ramond-Ramond Flux}, JHEP 9911 (1999) 003,
hep-th/9910205.}.  

A more ambitious
application would be to use the new formalism to construct a second-quantized
superstring field theory. Note that the physical state condition $Q V =0$
is easily generalized to the
non-linear equation of motion $Q V + V \wedge V =0$
where the $\wedge$-product is defined as in \ref\sft{E. Witten,
{\it Noncommutative Geometry and String Field Theory}, Nucl. Phys.
B268 (1986) 253.}. However, since the scattering amplitudes of
section 4 are only spacetime-supersymmetric when the states satisfy
the on-shell condition $QV=0$, it is unlikely that the action which
produces this equation of motion will be manifestly spacetime-supersymmetric.

Perhaps the most important unresolved issue 
is to prove the equivalence between this formalism and the
RNS formalism. In section 2, preliminary evidence for this
equivalence came from comparing the gluon vertex operator in the two
formalisms. Further evidence for this equivalence will now be shown
by considering the pure spinor formalism in the
``U(5) gauge'' defined by setting $u_{ab}= \t_{ab} =0$. 

In this U(5) gauge, the operator
$\oint dz \lambda^\a d_\a$ reduces to $\oint dz \g d_+$. If one compares 
with the U(5)-invariant formalism of \ufive, it is natural to
identify\foot{In the following discussion, 
$\gamma$ always refers to $\l^+$ and not to the RNS bosonic
ghost.}
$\g = e^{-\half(\rho-i\sigma)}$, or using
the field redefinition of \ufive\ to RNS worldsheet
variables,
$\g = c e^{-{3\over 2}\phi} \Sigma^+$
where $\Sigma^\a$ is the RNS spin field.
Since $d_+ = b \eta e^{{3\over 2} \phi} \Sigma_+$ using this
same field redefinition, 
one finds
$\oint dz\lambda^\a d_\a = \oint dz\eta$
in the U(5) gauge. So the restriction that 
$\oint dz \lambda^\a d_\a$ commutes with physical states maps in this gauge to
the restriction that physical states should
be independent of the $\xi$ zero mode.

Note that the unintegrated vertex operator for super-Yang-Mills,
$U=\l^\a A_\a$, reduces in the U(5) gauge to
\eqn\unin{
U = \g (A_+ + \t^a A_a + \half\t^a\t^b W_{ab} + {1\over{12}}
\e_{abcde}\t^a\t^b\t^c F^{de}
+{1\over {24}}\e_{abcde}\t^a \t^b\t^c\t^d \p^e W^+)}
where $A_a$ is the $\overline 5_{-1}$ component of the gluon gauge field and
$F^{ab}$ is the $10_2$ component of the gluon field strength.
Using the field redefinition of \ufive\
to write $U$ in terms of RNS variables
where
$\t^a = e^{\half\phi} \Sigma^a$,
one finds
\eqn\rnsunin{
U = c (e^{-{3\over 2}\phi}\Sigma^+
A_+ + e^{-\phi}\psi^a  A_a + \half e^{-\half\phi}\Sigma^{ab} W_{ab} 
+ {1\over{2}}
\psi_d\psi_e F^{de}
+e^{\half\phi}:\Sigma^- \psi_e: \p^e W^+)}
where $:\Sigma^- \psi_e:$ signifies $\lim_{y\to z} (y-z)^{-\half} \Sigma^-(y) 
\psi_e (z)$.
Except for the term proportional to $\p^e W^+$, all of the terms in
$U$ can be recognized as pieces of the RNS gluon and gluino vertex operators
in various pictures. 

Furthermore, if only the contribution from
$u_{ab}=0$ is included, the integration prescription of
\zeroq\ implies that 
\eqn\zerorns{\langle \l^\a \l^\b \l^\g
f_{\a\b\g}(z_r, k_r, \t)\rangle
\equiv ({\p\over{\p\t}})^5 f_{+++}(z_r, k_r, \t),}
i.e. $\langle \g^3 (\t)^5 \rangle =1$ where
$(\t)^5 = (5!)^{-1}\e_{abcde}\t^a\t^b\t^c\t^d\t^e$.
But under the field redefinition of \ufive, 
$\g^3(\t)^5$ maps to $c \p c \p^2 c e^{-2\phi}$, as expected from
$\langle \g^3 (\t)^5 \rangle =1$. 

So it appears reasonable that the U(5) gauge-fixed
version of the new formalism is equivalent to the RNS formalism with
a fixed choice of picture for the vertex operators. 
Removing the gauge-fixing condition, i.e.
integrating over the $u_{ab}$ and $\t_{ab}$ variables using the
prescription of \zeroq, might then be equivalent to summing over
the different possible pictures in the RNS formalism. It would
be very useful to be able to prove such an equivalence.

{\bf Acknowledgements:} I would like to thank CNPq grant 300256/94-9
for partial financial support.

\listrefs

\end